# Liquid phase production of graphene by exfoliation of graphite in surfactant/water solutions

Yenny Hernandez[1*], Mustafa Lotya[1*], Valeria Nicolosi[2], Fiona M Blighe[1] Sukanta De[1,3], Georg Duesberg[3,4] and Jonathan N Coleman[1,3**]

[1]*School of Physics, Trinity College Dublin, Dublin 2, Ireland*

[2]*Department of Materials, University of Oxford, Parks Road, Oxford OX1 3PH, UK*

[3]*CRANN, Trinity College Dublin, Dublin 2, Ireland*

[4]*School of Chemistry, Trinity College Dublin, Dublin 2, Ireland*

*These authors contributed equally to this paper.

**colemaj@tcd.ie

Abstract

We have demonstrated a method to disperse and exfoliate graphite to give graphene suspended in water-surfactant solutions. Optical characterisation of these suspensions allowed the partial optimisation of the dispersion process. Transmission electron microscopy showed the dispersed phase to consist of small graphitic flakes. More than 40% of these flakes had <5 layers with ~3% of flakes consisting of monolayers. These flakes are stabilised against reaggregation by Coulomb repulsion due to the adsorbed surfactant. However, the larger flakes tend to sediment out over ~6 weeks, leaving only small flakes dispersed. It is possible to form thin films by vacuum filtration of these dispersions. Raman and IR spectroscopic analysis of these films suggests the flakes to be largely free of defects and oxides. The deposited films are reasonably conductive and are semi-transparent. Further improvements may result in the development of cheap transparent conductors.

**Introduction**

The discovery of monolayer graphene in 2004[1] has led to the demonstration of a host of novel physical properties in this most exciting of nanomaterials[2]. Graphene is generally made by micromechanical cleavage, a process whereby monolayers are peeled from graphite crystals. However, this process has significant disadvantages in terms of yield and throughput. As such, there has been significant interest in the development of a large scale production method for graphene. In the long term, for many research areas the growth of graphene monolayers is by far the most desirable route. However, progress has been slow and in any case, this technique will be unsuitable for certain applications. Thus, in the medium term, the most promising route is the exfoliation of graphite in the liquid phase to give graphene-like materials. The most common technique has been the oxidisation and subsequent exfoliation of graphite to give graphene oxide.[3-7] However, this technique suffers from one significant disadvantage; the oxidisation process results in the formation of structural defects as evidenced by Raman spectroscopy[3, 6]. These defects alter the electronic structure of graphene so much as to render it semiconducting[8]. These defects are virtually impossible to remove completely; even after annealing at 1000°C, residual C=O and C-O bonds are observed by X-ray photoelectron spectroscopy[7]. Even mild chemical treatments, involving soaking in oleum, result in non-negligable oxidisation which requires annealing at 800°C to remove[9].

Recently, a significant breakthrough was made when two independent groups showed that graphite could be exfoliated in the liquid phase to give defect-free monolayer graphene[10, 11]. This phenomena relies on using special solvents whose surface energy is so well matched to that of graphene that exfoliation occurs freely[11]. However, this process is not without its drawbacks. These solvents are expensive and require special care when handling. In addition, they tend to have high boiling points, making it difficult to deposit individual monolayers on surfaces.

With these factors in mind, it is easy to see what is needed. We require a liquid phase process that results in the exfoliation of graphite to give graphene at reasonably high yield. The method should be non-oxidative and should not require high temperature processes or chemical post treatments. In addition it should be compatible with safe, user-friendly, low boiling-point solvents, preferably water.

In this paper we demonstrate such a method. We disperse graphite in surfactant-water solutions in a manner similar to surfactant aided nanotube dispersion[12-17]. By TEM analysis we demonstrate significant levels of exfoliation including the observation of a number of graphene monolayers. Raman and IR spectroscopy show the graphite/graphene to be defect free and un-oxidised. These dispersions can be vacuum filtered to make thin conductive films.

**Experimental procedure**

The graphite powder used in all experiments was purchased from Sigma Aldrich (product number 332461) and sieved through a 0.5 mm mesh to remove large particles. Sodium dodecylbenzene sulphonate (SDBS) was purchased from Sigma Aldrich (lot no. 065K2511) and used as provided. Stock solutions of SDBS of concentrations between 5 mg/ml and 10 mg/ml were prepared in Millipore water by stirring overnight. A typical sample was prepared by dispersing graphite in the desired SDBS concentration (25 ml sample volume in cylindrical vials) using 30 minutes of sonication in a low power sonic bath. The resulting dispersion was left to stand for approximately 24 hours to allow any unstable aggregates to form and then centrifuged for 90 minutes at 500 rpm. After centrifugation (CF), the top 15 ml of the dispersion was decanted by pipette and retained for use.

Sonication of the dispersions was carried out in a Branson 1510E-MT bath sonicator. Mild centrifugation was done using a Hettich Mikro 22R. Absorption measurements were taken with a Varian

Cary 6000i using quartz cuvettes. Sedimentation profiles were taken with a home-made apparatus using an array of synchronised pulsed lasers and photodiodes[18]. Annealing of some deposited films was carried out in a GERO Hochtemperaturöfen GmbH. Bright-field TEM images were taken with a Jeol 2100, operated at 200 kV. SEM analysis was carried out in a Hitachi S-4300 field emission SEM. Raman spectra were taken on a Horiba Jobin Yvon LabRAM-HR using a 100X objective lens with a 532 nm laser excitation. Attenuated total reflectance FTIR spectra were taken on a Perkin Elmer Spectrum 100. Thermogravimetric analysis was carried out using a Perkin Elmer Pyris 1 TGA in an oxygen atmosphere. The temperature was scanned from 25 to 900 ºC at 10 ºC /minute.

Zeta potential measurements were carried out on a Malvern Zetasizer Nano system with irradiation from a 633 nm He-Ne laser. The samples were injected in folded capillary cells, and the electrophoretic mobility ($\mu$) was measured using a combination of electrophoresis and laser Doppler velocimetry techniques. The electrophoretic mobility relates the drift velocity of a colloid (v) to the applied electric field (E); $v = \mu E$. All measurements were conducted at 25 °C and at the natural pH of the surfactant solution unless otherwise stated. The $\zeta$-potential can be calculated (in SI units) from the electrophoretic mobility using the Henry equation, incorporating the Smoluchowski approximation[17]: $\zeta = \eta \mu / \varepsilon$, where $\eta$ is the solution viscosity, $\varepsilon$ is the solution permittivity $\varepsilon = \varepsilon_r \varepsilon_0$. This equation is only rigorously valid for spherical particles. However, as it is known to predict $\zeta$-potential values for rod-like particles to within 20% of the true value we use it here for these disk-like systems accepting that some systematic error may be introduced.

**Results and discussion**

The absorption coefficient, α, which is related to the absorbance, A, through the Lambert-Beer law (A=αCl, where C is the concentration and l is the cell length), is an important parameter in characterising any dispersion. To accurately determine α, we prepared a dispersion (~400 ml) with initial graphite concentration, $C_{G,i}$ = 0.1 mg/ml, and SDBS concentration, $C_{SDBS}$ = 0.5 mg/ml. This was then centrifuged and decanted and the absorption spectrum measured (inset Figure 1). As expected for a quasi 2-dimensional material, this spectrum is flat and featureless[19] everywhere except below 280 nm where we observe a strong absorption band which scaled linearly with SDBS concentration but was independent of the graphite concentration; we attribute this band to the SDBS. A precisely measured volume of the dispersion was filtered under high vacuum onto an alumina membrane of known mass. The resulting compact film was washed with 1 L of water and dried overnight in a vacuum oven at room temperature. The mass of material in the filtered volume of stock dispersion was then determined using a microbalance. From TGA analysis (not shown) of the dried film, we found that 64 ± 5 % of the film was graphitic; the remainder was attributed to residual surfactant. This allowed us to determine the final concentration of the stock dispersion. A sample of the stock dispersion was then serially diluted with 0.5 mg/ml SDBS solution allowing the measurement of the absorbance per unit length (A/l) versus concentration of graphite (after centrifugation, $C_G$), as shown in Figure 1. A straight line fit through these points gives the absorption coefficient at 660 nm of α = 1390 ml mg$^{-1}$ m$^{-1}$ in reasonable agreement with the value measured for graphite/graphene in various solvents[11]. The non-zero intercept in Figure 1 is attributable to the A/l of residual SDBS in the dispersion (intercept of A/l=0.72 m$^{-1}$ compares with residual absorbance of A/l~0.5 m$^{-1}$ for SDBS at $C_{SDBS}$=0.5 mg/ml).

Using α for our dispersions, it is possible to determine $C_G$ for all subsequent samples. Thus, the fraction of graphite material remaining for any sample after centrifugation can be calculated from the ratio of dispersed graphite after CF to that before CF: $C_G/C_{G,i}$. Using this fraction-remaining as a gauge, the concentrations $C_{G,i}$ and $C_{SDBS}$ could be optimised. Holding $C_{SDBS}$ constant at a relatively high value of 10 mg/ml, $C_G$ was measured as a function of $C_{G,i}$ (Figure 2). Interestingly, we observe an empirical relationship of the form: $C_G = 0.01\sqrt{C_{G,i}}$. The largest fraction remaining was ~3wt% at $C_{G,i}$ = 0.1 mg/ml (top inset, Figure 2). This graphite concentration was then fixed and $C_{SDBS}$ varied. Measurement of the fraction remaining showed a broad peak (lower inset, Figure 2), similar to those observed for nanotube-surfactant dispersions[17], with reasonably high quantities of graphite remaining for $C_{G,i}$ between 0.5 mg/ml to 1 mg/ml. The fall-off in dispersed graphite below $C_{SDBS}$~0.5 mg/ml is reminiscent of the destabilisation of nanotube dispersions as the surfactant concentration is reduced below the critical micelle concentration[17, 20] (~0.7 mg/ml for SDBS[21]). Keeping the concentration of surfactant to a minimum is desirable for many potential applications, so, all subsequent experiments were performed on standard dispersions with $C_{SDBS}$ = 0.5 mg/ml and $C_{G,i}$ = 0.1 mg/ml. (NB the fraction remaining in the experiment described in Figure 1 was much smaller than would be expected from the data shown in Figure 2. This is due to the fact that in the former experiment a much larger volume was used resulting in less efficient sonication.)

At this point we know we are dispersing graphite but not in what form. To further characterise the dispersions, we conducted TEM analysis on our standard dispersion. TEM samples were prepared by pipetting a few milliliters of this dispersion onto holey carbon mesh grids (400 mesh). TEM analysis revealed a large quantity of flakes of different types as shown in Figure 3. A small quantity of monolayer graphene flakes were observed (Figure 3A). A larger proportion of flakes were few-layer graphene, including some bilayers and trilayers as shown in Figure 3B and C. In addition, a number of rather disordered flakes with many layers, similar to the one in Figure 3D were observed. The disorder suggests that these flakes formed by reaggregation of smaller flakes. Finally, a very small number (2) of very large flakes were observed (Figure 3E). It can be shown that these are graphite by the observation of thin multilayers protruding from their edges (Figure 3E, Inset). Note that while these large flakes are rare when counted by number, they are expected to contribute disproportionally by mass. It is possible to estimate the number of layers per flake for every flake observed. This data is illustrated in the histogram for the standard dispersion in Figure 4A (the very large flakes are ignored in this histogram). These statistics show a reasonable population of few-layer graphene. For example ~43% of flakes has <5 layers. More importantly, ~3% of the flakes were monolayer graphene. While this value is considerably smaller than that observed for graphene/solvent dispersions[11], working in aqueous systems brings its own advantages. In general, the majority of these few-layer flakes had lateral dimensions of ~1μm. Thicker flakes, with more than a few graphene layers per flake, were larger, ranging up to 3 µm in diameter.

The sediment remaining after centrifugation can be recycled to improve the overall yield of graphene exfoliation. The sediment was dried and fresh 0.5 mg/ml SDBS solution was added. This sediment dispersion was then processed in the same manner as the original dispersion and TEM analysis carried out. In this case, we also observed the presence of isolated monolayer graphene in about 3% of cases (Figure 3F). In addition, the flake thickness distribution shifted towards thinner flakes with large quantities of bilayers and trilayers; 67% of flakes observed had <5 layers (Figure 4B). Notably, there were no large flakes with greater than 10 layers observed, indicating that the reprocessing of recycled sediment gives better exfoliation than processing of the original sieved graphite. We suggest that the second sonication breaks up the already partially exfoliated chunks of graphite into even smaller pieces from which exfoliation occurs more easily. As such, it is unlikely that simply doubling the sonication time would yield equivalent results.

The zeta potential is useful parameter we can use to characterise our dispersions. SDBS is an ionic surfactant that is expected to adsorb onto the graphene flakes and impart an effective charge. We expect that the dispersions will be stabilised by electrostatic repulsion between surfactant-coated graphene flakes. This mechanism has allowed the successful dispersion of carbon nanotubes in a range of surfactants.[16, 17, 22, 23] The zeta potential is the potential at the interface between the adsorbed surfactant molecular ions and the diffuse region of mobile counterions. As such it is a measure of the electrostatic repulsion between surfactant coated flakes. Our dispersions are in aqueous media with free $Na^+$ counterions and so have high ionic conductivity. In addition, we make the (crude) approximation that our planar graphene flakes can be treated as spherical particles whilst in dispersion. Hence, we apply the Smoluchowski approximation in our measurements[24]; this is in line with previous work on carbon nanotube dispersions in SDBS.[17, 22] The natural pH of our dispersions was 7.4, which matches a literature value for SDBS stabilised carbon nanotube dispersions.[22]

We observed a zeta potential distribution for a fresh graphite/graphene dispersion centred at -44 mV (Figure 5A). The shoulder at -76 mV is probably due to free surfactant, as it matches well to the position of the zeta spectrum of a 0.5 mg/ml SDBS solution at -71 mV (Figure 5A). For fresh graphite/graphene, the peak zeta potential of -44 mV is well below the accepted value for colloidal stability of -15mV, indicating that reaggregation should be minimised. For comparison, the zeta spectrum of a 6 week old graphene/graphite dispersion is also shown. This spectrum is peaked at -78 mV with a shoulder at -103 mV. We suggest the peak is due to unbound surfactant while the shoulder is due to surfactant coated graphite/graphene flakes. That the zeta potential has shifted to more negative values over six weeks strongly suggests that the electrophotetic mobility, µ, has increased in magnitude. One explanation for this could be a reduction in mean flake size which may increase the electrophoretic mobility in non-spherical samples. The origin of such a size reduction will be discussed below.

The pH of the fresh dispersion was varied by addition of HCl and NaOH with the results given in Figure 5A (inset). There is a trend towards more negative zeta potential values as the pH is raised; this suggests that inter-particle repulsions are increased as more negative $OH^-$ charges are added to the flakes. For acidic dispersions at lower pH values a less negative zeta potential is found, consistent with charge neutralisation and destabilisation of the system. The zeta potential vs pH trend is in line with trends reported for graphene oxide and reduced graphene oxide colloids.[4] By lowering the pH, the zeta potential approaches the limit of stability in our system but it does not pass through the isoelectric point. This maybe due to very high surface coverage of graphene flakes by SDBS ionic molecules and perhaps also due to a buffer-like action by the free surfactant in the dispersion.

To determine the temporal stability of these dispersions, we conducted sedimentation experiments on a centrifuged, decanted dispersion ($C_G$=0.006 mg/ml, $C_{SDBS}$=0.5 mg/ml). The optical absorbance of the sample at 650 nm was monitored as a function of time as shown in Figure 5B. The measured absorbance fell steadily, indicating sedimentation of approximately two thirds of the material over a considerable period of time. A bi-exponential function could be fitted to the profile, indicating one stable and two sedimenting components[18]. The fit parameters indicate that 35wt% of the sample is stable over the timeframe of 35 days. We attribute this component to small flakes. Of the rest, 19 wt% of the flakes fall out rapidly, with a time constant of 21.5 hrs while a further 46 wt% fall out over longer time scales (time constant ~208 hrs). As the time constant is related to the dimensions of the sedimenting object[18], we can attribute the slowly and rapidly sediment objects to medium and large sized flakes respectively. We suggest that the large flakes are fragments of graphite that inadvertently remained in the dispersion after decantation and which we can identify with the type of flake observed in Figure 4E. We identify the medium sized flakes as those objects represented at the right side of the histogram in Figure 4A. TEM analysis of the 6 week old sample used for zeta measurements showed only small flakes remain; these were typically few-layer graphene flakes less than 500 nm in diameter. This confirms both that medium to large flakes are unstable and sediment out over 6 weeks (~1000 hrs) and that the increase in $|\zeta|$ is due to an increase in $|\mu|$ caused by the reduction in flake size over time.

To examine the potential uses of aqueous graphene dispersions, films were cast onto porous membranes by vacuum filtration. Film preparation was carried out immediately after CF using nitrocellulose membranes (pore size 25 nm) or alumina membranes (pore size 20 nm) supported on a fritted glass holder. These films were dried overnight in a room temperature vacuum oven at $\sim 1\times 10^{-3}$ mbar to remove the water. Figure 6 shows SEM and optical images of a typical film (the segment of the film used for SEM was coated with 10 to 20 nm of gold). It can be seen from the SEM image that many of the flakes are small with diameters ~1 µm. In addition, there are some large flakes ~5 µm in diameter which we associate with the flake shown in Figure 3E. In contrast to films cast from solvents[11], the flakes lie flat on top of each other, suggesting the possibility of good electrical contact between flakes. The small flakes are not visible in the optical image, appearing as a constant background. However, the large flakes *are* apparent, appearing as bright regions. Significant quantities of these large flakes are present.

The deposited films were further characterised by Raman spectroscopy. Examples of typical film spectra are given in Figure 7, alongside a spectrum for the starting graphite powder (these spectra were normalised to the intensity of the *G*-band at 1582 cm$^{-1}$). Spectra of graphitic materials are characterised by a *D*-band (1350 cm$^{-1}$), a *G*-band (1582 cm$^{-1}$) and a *2D*-band (2700 cm$^{-1}$). The studied film had been deposited on an alumina membrane and rinsed with 17.5 ml of water before drying. As was the case in the film shown in Figure 6, this film consists of large flakes (diameter~3-6µm) embedded in a matrix of small flakes (diameter~1µm). Shown in Figure 7 are Raman spectra collected by focusing the laser spot both on the region of small flakes and on a large flake. Like the starting graphite powder, no *D*-band (1350 cm$^{-1}$) is observed in the spectrum associated with the large flake. This shows that the dispersion process does not result in the formation of defects on the graphitic basal plane. In addition, the *2D*-band of this large flake strongly resembles the *2D*-band for graphite. This indicates that this flake is relatively thick with >5 graphene layers[25]. The relatively large diameter and thickness of such flakes allows us to associate them with the large flakes observed in Figure 3E and those which rapidly sediment out of the dispersions measured in Figure 5B. In the case of the spectrum associated with the region of small flakes, a *D*-band is observed. We stress that this *D*-band is both narrower and less intense than that reported in literature for graphene oxide and for reduced graphene oxide.[3, 6] We attribute this feature to edge effects as the Raman excitation beam spot size of ~ 2 µm is larger than most of the flakes in the deposited film. The relatively low *D*-band intensity observed for the small flakes coupled with the complete absence of a *D*-band for the bigger flakes strongly suggests that the films we are producing are composed of very low defect material. In addition, by comparison to literature[25], the shape of the *2D*-band observed for the small flakes is characteristic of thin flakes composed of less than five graphene layers. This shows that while re-aggregation undoubtedly occurs during filtration, the degree of re-aggregation is limited.

Attenuated total reflectance (ATR) FTIR spectra of deposited films were also measured as a function of washing regime (Figure 8A), along with reference spectra for SDBS powder and the alumina membrane (Figure 8B). These spectra show only very small features at ~ 1100 cm$^{-1}$ and ~ 2900 cm$^{-1}$. By comparison with the reference spectra, it is clear that these features are attributable to residual surfactant trapped in the film. A key feature of the spectra in Figure 8A is the complete absence of peaks associated with C–OH ( ~1340 cm$^{-1}$) and –COOH ( ~1710 to 1720 cm$^{-1}$) groups.[4, 26-28] Our spectra are in contrast to those in the literature for films made from reduced graphene oxide[4, 28] or chemically derived graphene[9]. This is further evidence that our exfoliation technique does not chemically functionalise the graphene/graphite and that our films are composed of largely defect free material.

In order to test the optical and electrical properties of these films, we measured the transparency (632 nm) and sheet resistance of a number of vacuum deposited films (nominal thickness ~ 30nm). As-deposited films typically had transmittance of ~62% coupled with sheet resistance of ~970 kΩ/ . This corresponds to a DC conductivity of 35 S/m. The low value is probably attributable to the presence of large quantities of residual surfactant. As discussed above, up to 36wt% of filtered films is residual surfactant which can be difficult to remove by washing. We attempted to remove the surfactants by annealing @ 250°C in Ar/N$_2$ for 2 hrs prior to re-measuring the transmittance and sheet resistance. After annealing the transparency was unchanged while the sheet resistance had fallen to 22.5 kΩ/ , consistent with a nominal DC conductivity of 1500 S/m. This value is significantly lower than that recently measured for similar films prepared from N-methyl-pyrrolidone based dispersions (~6,500 S/m). In addition, films of reduced graphene oxide have displayed conductivities ranging from 7,200 S/m[4] to 10,000 S/m[7]. In comparison, graphene dispersed in dimethyl-acetamide has been spray-cast into films with conductivities as high as $10^5$ S/m.[10] Thus the presence of residual surfactant severely impedes the electrical properties of our films. However, we believe that the combination of aqueous environment and lack of defects give our dispersion / exfoliation method great potential. Complete removal of surfactant may result in a material which can challenge nanotubes as an indium tin oxide replacement material. Future work will focus on removal of residual surfactant from films, the maximisation of electrical conductivity and the deposition of individual monolayers.

**Conclusion**

We have developed a method to disperse graphite in surfactant-water solutions with the aid of ultrasound. This results in large scale exfoliation to give large quantities of multilayer graphene with <5 layers and smaller quantities of monolayer graphene. The exfoliated flakes are stabilised against reaggregation by Coulomb repulsion due to the adsorbed surfactant molecular ions. The dispersions are reasonably stable with larger flakes sedimenting out over ~6 weeks. These dispersions can be used to form films by vacuum filtration. Characterisation of the films by Raman and IR spectroscopy suggest the absence of defects or oxides on the graphene basal plane. These films are reasonably conductive and can be made semi-transparent. It is anticipated that their properties can be significantly enhanced by improved surfactant removal.

**Acknowledgements**

The authors would like to acknowledge IRCSET and Science Foundation Ireland, through the Principle Investigator scheme for financial support.

**Figures**

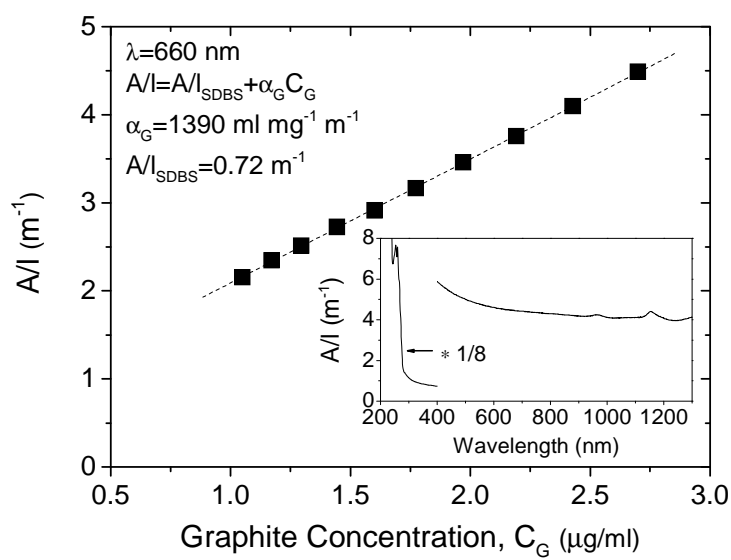

Hernandez et al

Figure 1: Absorbance per unit length (λ=660 nm) as a function of graphite concentration (after centrifugation) for an SDBS concentration, $C_{SDBS}$=0.5 mg/ml. Graphite concentration before centrifugation was $C_{G,i}$=0.1 mg/ml. NB, the curve does not go through the origin due to the presence of a residual SDBS absorbance. (Intercept of A/l=0.72 m$^{-1}$ compares with residual absorbance of A/l~0.5 m$^{-1}$ for SDBS at $C_{SDBS}$=0.5 mg/ml). Inset: Absorption spectrum for a sample with $C_{SDBS}$=0.5 mg/ml and $C_G$=0.0027 mg/ml. The portion below 400 nm is dominated by the surfactant absorption and has been scaled by a factor of 1/8 for clarity. The portion above 400 nm is dominated by graphene/graphite with some residual SDBS absorption.

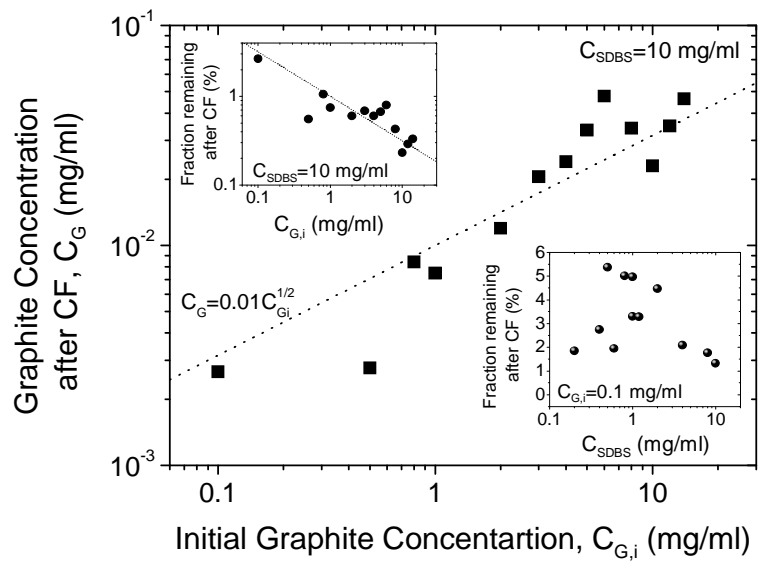

Figure 2: Graphite concentration after centrifugation (CF) as a function of starting graphite concentration ($C_{SDBS}$=10 mg/ml). Upper inset: The same data represented as the fraction of graphite remaining after CF. Lower inset: Fraction of graphite after centrifugation as a function of SDBS concentration ($C_{G,i}$=0.1 mg/ml).

Hernandez et al

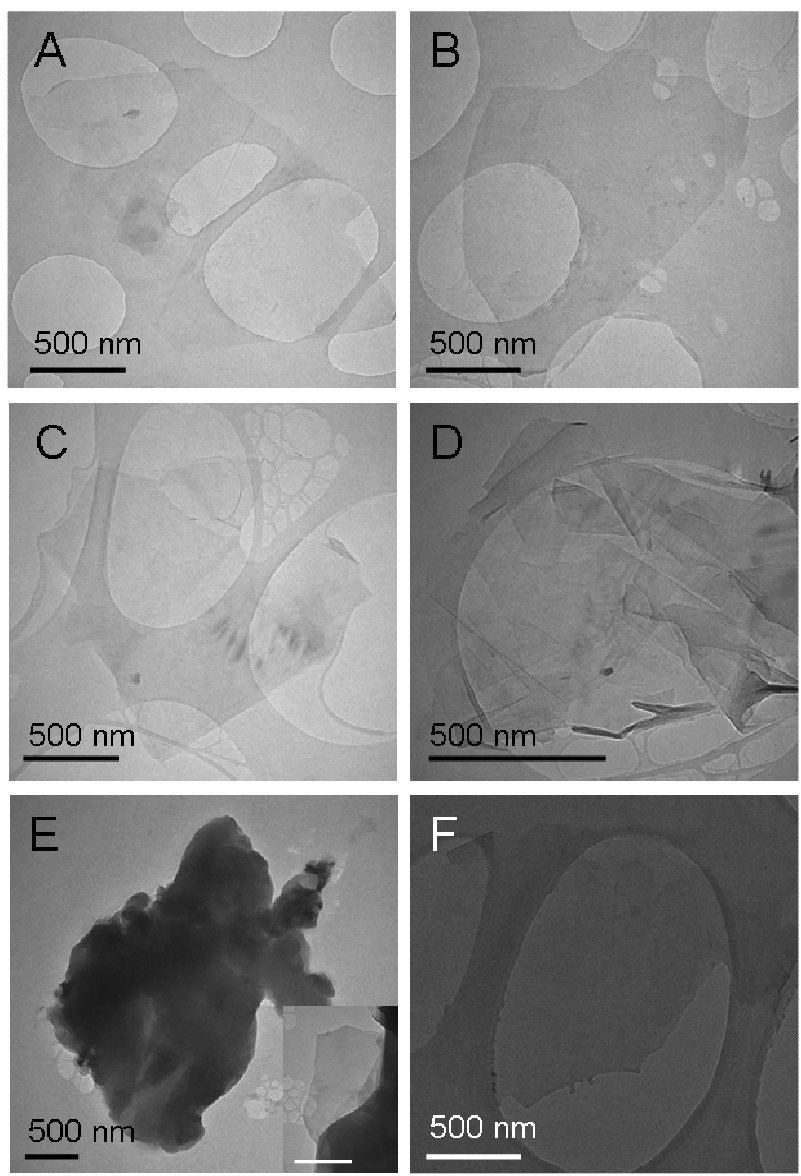

Figure 3: Selected TEM images of flakes prepared by surfactant processing. A) a monolayer (albeit with a small piece of square debris close to its left hand edge). B) A bilayer. C) A trialyer. D) A disordered multilayer. E) A very large flake. Inset: A closeup of an edge of a very large flake showing a small multilayer graphene flake protruding (scale bar 500 nm). F) A monolayer form a sample prepared by sediment recycling.

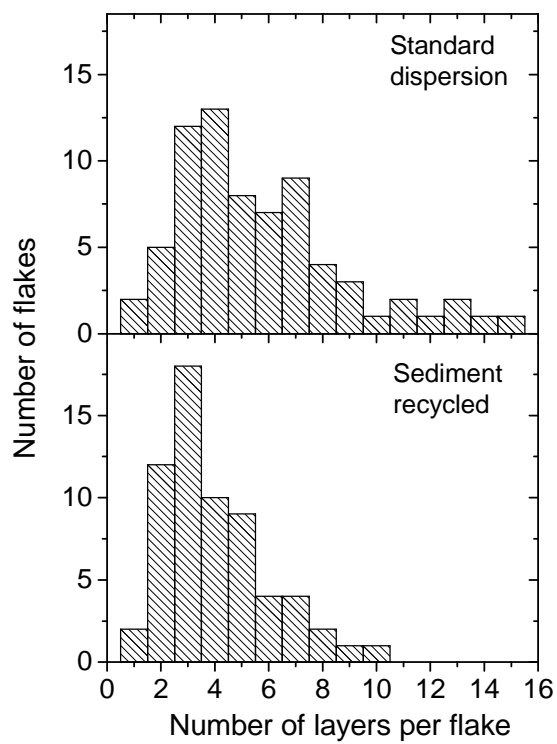

Figure 4: Histogram of number of layers per flake for dispersions from original sieved graphite and from recycled sediment. This histogram does not include the two very large flakes of the type shown in figure 3E.

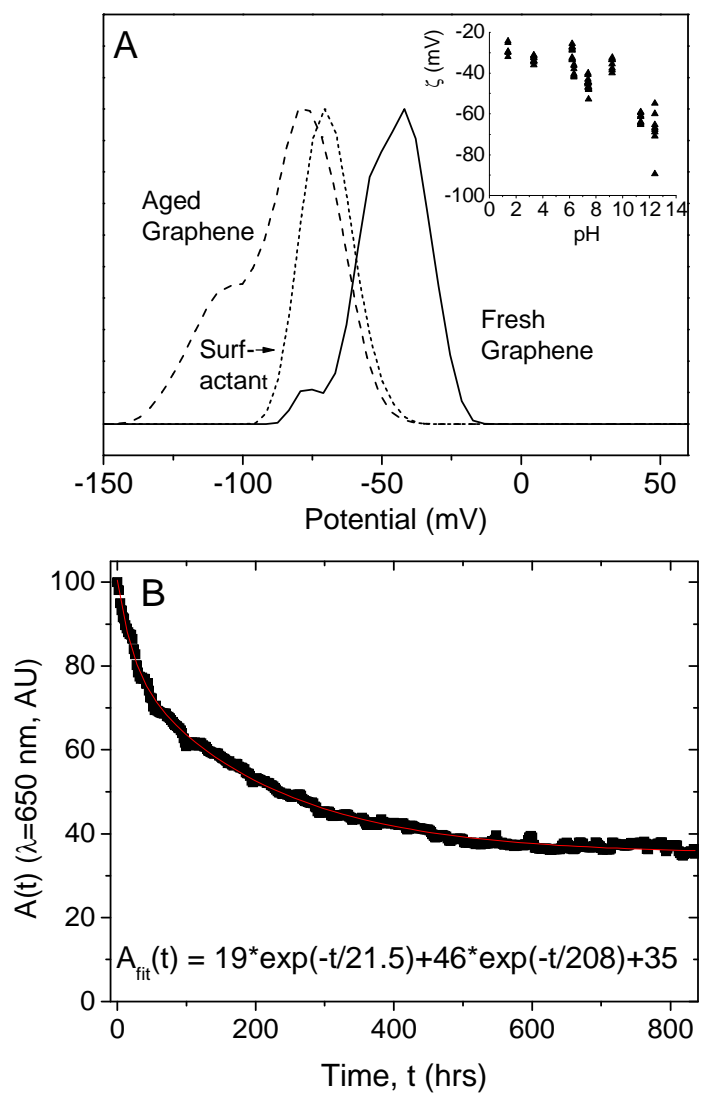

Figure 5: A) Zeta spectra for a fresh graphene-SDBS dispersion ($C_{SDBS}$=0.5mg/ml, $C_G$=0.006mg/ml), an SDBS dispersion ($C_{SDBS}$=0.5mg/ml) and an aged (6 week old) graphene-SDBS dispersion ($C_{SDBS}$=0.5mg/ml, $C_G$=0.0002mg/ml). NB the aged sample had a reduced $C_G$ due to sedimentation over the course of 6 weeks. Inset: Zeta potential as a function of pH for SDBS-graphene dispersions ($C_{SDBS}$=0.5mg/ml, $C_G$=0.005mg/ml). The natural pH of the as prepared graphene-SDBS dispersion was 7.4 and the pH was varied by addition of HCl or NaOH solution. B) Absorbance ($\lambda$=650 nm) as a function of time for a $C_G$=0.006 mg/ml, $C_{SDBS}$=0.5 mg/ml sample. The curve has been fitted to a double exponential decay with the fit constants shown in the annotation.

Hernandez et al

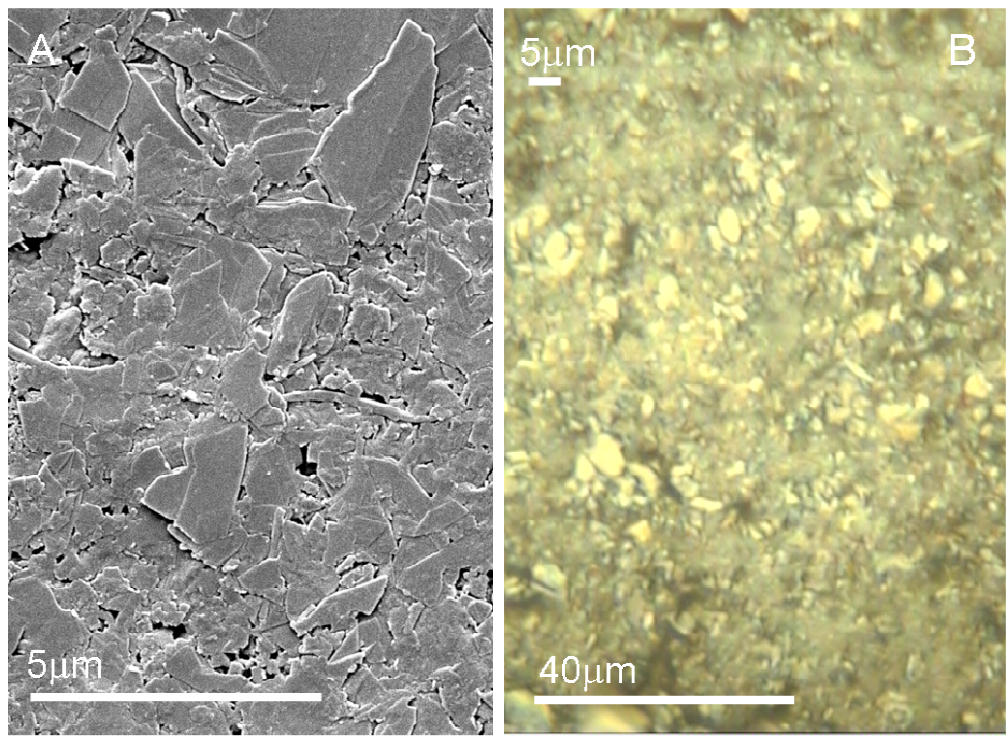

Figure 6: A) SEM and B) optical images of the surface of a graphene film. This film was ~150 nm thick and had been deposited on a cellulose membrane by filtration from an SDBS based dispersion ($C_{SDBS}$=0.5mg/ml, $C_G$=0.003mg/ml). This film was not rinsed and was dried under vacuum at room temperature.

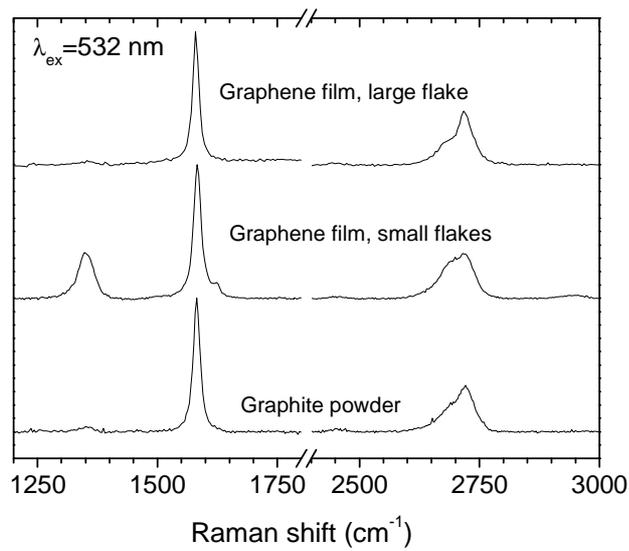

Figure 7: Raman spectrum of a graphene film (thickness ~300nm) deposited on an alumina membrane by filtration from an SDBS based dispersion ($C_{SDBS}$=0.5mg/ml, $C_G$=0.005mg/ml) and rinsed with 17.5 ml of water. Spectra associated with both large flakes (diameter~3-6μm, top) and small flakes (diameter~1μm, middle) are shown. For comparison, a spectrum collected from the starting graphite powder is included (bottom).

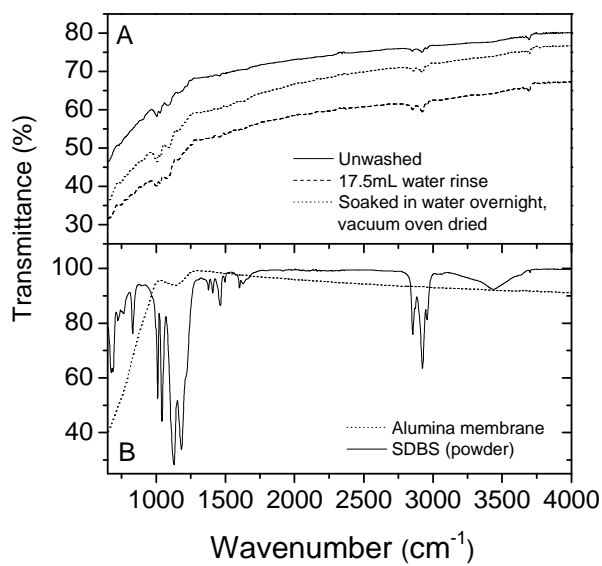

Figure 8: ATR-FTIR spectra of materials used in this study. A) Spectra of three graphene films with different washing regimes. The films were ~300nm thick and were deposited on alumina by vacuum filtration from an SDBS based dispersion ($C_{SDBS}$=0.5mg/ml, $C_G$=0.005mg/ml). B) Control spectra of SDBS powder and the alumina filter membrane used to prepare the graphene films.